\documentclass[12pt]{iopart}

\usepackage{graphicx}
\usepackage{graphics}
\usepackage[usenames]{color}
\begin{document}
\title
[Entanglement Spectrum in Cluster Dynamical Mean-Field Theory]{Entanglement Spectrum in Cluster Dynamical Mean-Field Theory}

\author{Masafumi Udagawa and Yukitoshi Motome}

\address{Department of Applied Physics, University of Tokyo, Hongo 7-3-1, Bunkyo-ku, Tokyo 113-8656, Japan}
\ead{udagawa@ap.t.u-tokyo.ac.jp}
\begin{abstract}
We study the entanglement spectrum of the Hubbard model at half filling on a kagome lattice. The entanglement spectrum is defined by
the set of eigenvalues of reduced thermal density matrix, 
which is naturally obtained in the framework of the dynamical mean-field theory. 
Adopting the cluster dynamical mean-field theory combined with continuous-time auxiliary-field Monte Carlo method, we calculate the entanglement spectrum for a three-site triangular cluster in the kagome Hubbard model. 
We find that the results at the three-particle sector
well captures the qualitative nature of the system. In particular, the eigenvalue of the reduced density matrix, corresponding to the
chiral degrees of freedom, exhibits characteristic temperature scale $T_{\rm chiral}$, 
below which
a metallic state with large quasiparticle mass is stabilized. The entanglement spectra at different particle number sectors also
exhibit characteristic changes around $T_{\rm chiral}$, implying the development of inter-triangular ferromagnetic correlations 
in the correlated metallic regime.

\end{abstract}

\maketitle

\section{Introduction}
It is a central issue of condensed matter physics to describe the nature of strongly correlated electron systems both qualitatively and quantitatively. Dynamical mean-field theory (DMFT)~\cite{Georges96} provides a promising theoretical tool for this purpose.
DMFT bridges the gap between two solvable cases, the atomic limit and the free fermion limit, and successfully describes the properties 
of correlated electron systems. In particular, correlation-driven metal-insulator transition (MIT) is successfully explained by DMFT. 
In addition to the success in describing various thermodynamic quantities in a quantitative way, DMFT also provides a qualitative
understanding 
of MIT in terms of insulating Hubbard bands and a renormalized conduction band.
These aspects of MIT can be qualitatively captured by focusing on a simple observable quantity, a doublon density, i.e., the fraction of sites where spin-up and spin-down states are simultaneously occupied. The temperature dependence of this quantity 
is easily computable in the scheme of DMFT and it, indeed, gives crucial insights into the nature of MIT~\cite{Georges93}.

While DMFT is quite useful to understand MIT in a simple single-orbital correlated model, however, realistic compounds
have complex degrees of freedom composed of spin and orbitals and their spatial correlations, which lead to rich behavior 
in MIT. 
Correspondingly, the cluster extension of DMFT (CDMFT)~\cite{Kotliar01} has been used to explore this fertile field, by
taking into account 
spin, orbital, and intersite degrees of freedom, in a quantitative way.
However, the complexity of the system inevitably makes it difficult to achieve qualitative understanding.
It is then, desirable to set up a measure to characterize these systems in a qualitative way, i.e., the counterpart of doublon density
in these complex systems. For this purpose, in this letter, we propose the entanglement spectrum as a measure to qualitatively describe
the systems composed of multiple degrees of freedom. 

This quantity can be computed with a small numerical cost, and gives account to qualitative nature of the system.
In contrast to the usual terminology of entanglement spectrum, it is defined 
from the 
reduced thermal density matrix.
Accordingly, standard properties required for the entanglement spectrum, e.g., the area law of entanglement entropy, 
will not be satisfied. Nonetheless, we show that this quantity will be quite useful to disentangle the 
complexity of correlated systems
with multiple degrees of freedom. 

This letter is organized as follows. In section~\ref{modelmethod}, we introduce CDMFT, and especially the algorithm 
to compute the entanglement spectrum. In section~\ref{results}, we show the results of entanglement spectrum, and discuss
how this quantity is associated with the actual behavior of the systems, by taking the Hubbard model on the kagome lattice as an example.
Section~\ref{sec:summary} is devoted to the discussion and summary.

\section{Model and Method}
\label{modelmethod}
\subsection{Model}
\begin{figure}[t]
\begin{center}
\includegraphics[width=0.9\textwidth]{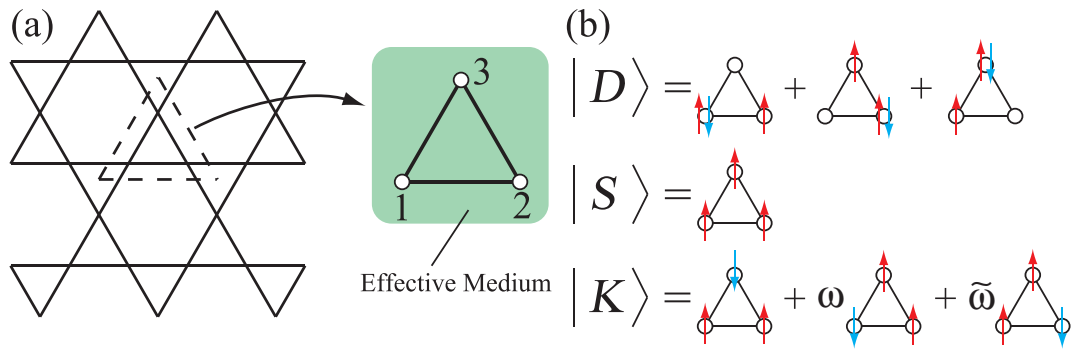}
\end{center}
\caption{\label{schematics} 
(color online). (a) Schematic picture of the mapping
of the kagome lattice to three-site cluster in the
CDMFT calculations. The three-site density matrix is calculated
for the numbered three sites. 
(b)
Schematic pictures of the doublon, spin-polarized, and chiral
states.}
\end{figure}

As a prototypical model of correlated electron systems composed of multiple degrees of freedom, we consider
the single-orbital Hubbard model on the kagome lattice. 
The Hamiltonian is given in the standard notation as 
\begin{eqnarray}
\mathcal{H}= -t \sum_{\langle ij \rangle,\sigma}
(a_{i\sigma}^\dagger a_{j\sigma} + {\rm {h.c.}})
+ U \sum_i n_{i\uparrow} n_{i\downarrow},
\label{eq:H}
\label{HubbardHamiltonian}
\end{eqnarray}
where the sum of $\langle ij \rangle$ is taken over the nearest-neighbor sites on the kagome lattice (Fig.~\ref{schematics}). 
Hereafter we set $t=1$ and the Boltzmann constant $k_{\rm B}=1$, and fix the electron density at half filling.
This model has been extensively studied as a minimal model with electron correlation and geometrical frustration~\cite{Udagawa10,Imai03,Bulut05,Ohashi06,Bernhard07,Furukawa10,Udagawa10_2}.

The kagome lattice has three sites in the unit cell, which constitute 
the internal degrees of freedom.
The interplay between the intra-cell degrees of freedom and electron correlation leads to nontrivial
heavy-fermion behavior in the metallic region, as we will show later.

\subsection{Cluster DMFT}
The model (\ref{HubbardHamiltonian}) is not easily tractable with a standard unbiased numerical method, e.g., exact diagonalization or quantum Monte Carlo simulation, due to the exponentially large numerical cost or notorious negative sign problem.
We here adopt the cluster dynamical mean-field theory (CDMFT)~\cite{Georges96,Kotliar01} as an approximate method to take into account 
the electron correlation. In 
CDMFT, the Hubbard model (\ref{eq:H}) on the original infinite lattice 
is mapped to the effective Anderson model with interacting impurity sites 
on an $N_{\rm c}$-site cluster embedded in the effective medium. In this study, we mainly focus on the case of $N_{\rm c}=3$, namely, we 
adopt the triangle unit cell as an impurity cluster, as shown in Fig.~\ref{schematics}(a).
In order to solve the effective Anderson model, we adopt the continuous-time auxiliary-field quantum Monte Carlo method~\cite{Gull08}.
This method is based on the perturbation expansion in terms of the interaction parameter 
$U$, and is particularly suitable for the calculation on a larger-size cluster.

\subsection{Entanglement Spectrum}
The doublon density, defined by $D\equiv\langle n_{i\uparrow}n_{i\downarrow}\rangle$, is a standard measure to characterize the effect of electron correlation. 
$D$ corresponds to the fraction of sites where the up-spin and down-spin states are 
simultaneously occupied. 
It shows a characteristic temperature change, accompanying 
a discontinuity between metallic and insulating states. This quantity is, however, not enough to 
characterize the complicated energy scales in the correlated electron systems with multiple degrees of freedom.

To generalize $D$ for the systems with multiple degrees of freedom, we focus on the aspect of doublon density as an eigenvalue
of the reduced density matrix. In DMFT, the whole system is divided into the single target site and the environment.  
By tracing out the environment degrees of freedom, 
the reduced density matrix $\hat{\rho}$ is constructed for the target site, as
\begin{eqnarray}
\hat{\rho} = \rho_{\uparrow\downarrow}|\uparrow\downarrow\rangle\langle\uparrow\downarrow| + \rho_{\uparrow}|\uparrow\rangle\langle\uparrow| + \rho_{\downarrow}|\downarrow\rangle\langle\downarrow| + \rho_{0}|0\rangle\langle0|.
\label{reduced_dmatrix_single}
\end{eqnarray}
Here, $|\uparrow\downarrow\rangle=a^{\dag}_{\uparrow}a^{\dag}_{\downarrow}|0\rangle$, $|s\rangle=a^{\dag}_{s}|0\rangle$, and $|0\rangle$ are a doubly-occupied state, a
singly-occupied state with spin $s$ electron, and the empty state, respectively.
In equation~(\ref{reduced_dmatrix_single}), $\hat{\rho}$ takes a diagonalized form by choosing the particle number $N_p$ and spin
quantum number 
to label the bases, in the systems with conserved particle number and spin. 
In this representation,
the doublon density $D$ is nothing but the coefficient $\rho_{\uparrow\downarrow}$, i.e., the eigenvalue of $\hat{\rho}$ in the $N_p=2$ sector, in equation~(\ref{reduced_dmatrix_single}). 

\subsection{CDMFT Treatment}
Now, let us consider the CDMFT 
calculations of the entanglement spectrum for the systems composed of multiple degrees of freedom.
The cluster impurity mapping of CDMFT divides the system into the (A) cluster sites and (B) environment, which
gives a natural definition of reduced density matrix, $\hat{\rho}_{\rm A}\equiv{\rm Tr}_{\rm B}\hat{\rho}_{\rm thermal}$. Here $\hat\rho_{\rm thermal}\equiv\exp{(-\beta\mathcal{H})}/
{\rm Tr}_{\rm A} {\rm Tr}_{\rm B} \exp{(-\beta\mathcal{H})}$ is the equilibrium thermal density matrix, and the trace 
${\rm Tr}_{\rm A(B)}$ is taken over the cluster (environment) degrees of freedom.
The reduced density matrix $\hat{\rho}_{\rm A}$ can be expanded with the $4^{N_{\rm c}}$ basis states $|\psi_p\rangle$ within the cluster sites,
as $\hat{\rho}_{\rm A} = \rho_{{\rm A}, pq}|\psi_p\rangle\langle\psi_q|$. $\hat{\rho}_{\rm A}$ can be diagonalized to give $4^{N_{\rm c}}$ eigenvalues, $\{\rho_{{\rm A},pq}\}$,
which constitute the entanglement spectrum.

In order to obtain each coefficient $\rho_{{\rm A}, pq}$ in the CDMFT formalism, it is convenient to adopt the second 
quantized representation.
We choose the basis to diagonalize the particle number and the $z$ component of spin: 
\begin{eqnarray}
|\phi\{m_{is}\}\rangle=(a^{\dag}_{1\uparrow})^{m_{1\uparrow}}(a^{\dag}_{1\downarrow})^{m_{1\downarrow}}\cdots(a^{\dag}_{N_{\rm c}\uparrow})^{m_{N_{\rm c}\uparrow}}(a^{\dag}_{N_{\rm c}\downarrow})^{m_{N_{\rm c}\downarrow}}|0\rangle. 
\end{eqnarray}
Here, $m_{is}=0$ or $1$ is the number of electron at site $i$ and spin $s$. The projection operator $|\psi_p\rangle\langle\psi_p|$ can be simply expressed as
\begin{eqnarray}
|\phi\{m_{is}\}\rangle\langle\phi\{m_{is}\}| = \hat{f}_{1\uparrow}[m_{1\uparrow}]\hat{f}_{1\downarrow}[m_{1\downarrow}]\cdots\hat{f}_{N_{\rm c}\uparrow}[m_{N_{\rm c}\uparrow}]\hat{f}_{N_{\rm c}\downarrow}[m_{N_{\rm c}\downarrow}],
 \label{density_diagonal}
\end{eqnarray}
where $\hat{f}_{is}[m] = 1-n_{is}$ $(n_{is})$ for $m=0$ $(1)$. Accordingly, the diagonal components of $\hat{\rho}_{\rm A}$, $\rho_{{\rm A}, pp}$, can be obtained as the expectation value of the corresponding operator $|\phi\{m_{is}\}\rangle\langle\phi\{m_{is}\}|$. 

The calculation of the off-diagonal components of $\hat{\rho}_{\rm A}$ is slightly more involved. To depict this, suppose the three-site cluster ($N_{\rm c}=3$),
and consider 
an off-diagonal component, $|\uparrow, 0, 0\rangle\langle0, \uparrow, 0|$. Here, we denote $|\uparrow\downarrow, \uparrow, 0\rangle\equiv a_{1\uparrow}^{\dag}a_{1\downarrow}^{\dag}a_{2\uparrow}^{\dag}|0\rangle$, and so on. 
Using the expressions, we can write down the creation operator as
\begin{eqnarray}
a^{\dag}_{1\uparrow} = \sum_{k_2,k_3}(|\uparrow, k_2, k_3\rangle\langle0, k_2, k_3| + |\uparrow\downarrow, k_2, k_3\rangle\langle\downarrow, k_2, k_3|).
\end{eqnarray}
Here, the summation of $k_i$ is taken over the four possible states at site $i$: $|0\rangle, |\uparrow\rangle, |\downarrow\rangle$ and $|\uparrow\downarrow\rangle$.
Then, we obtain 
\begin{eqnarray}
|\uparrow, 0, 0\rangle\langle0, \uparrow, 0| = a^{\dag}_{1\uparrow}a_{2\uparrow} |0, \uparrow, 0\rangle\langle0, \uparrow, 0|. 
\end{eqnarray}
Combining 
this type of expressions and the projection operator given in equation~(\ref{density_diagonal}), the off-diagonal components can also be expressed in the second quantized form.
Accordingly, all the components of the reduced density matrix, $\{\rho_{{\rm A}, pq}\}$, can be expressed as the product of fermion creation and annihilation operators. 
Hence, standard theoretical techniques, such as Wick's theorem, can be used for their computation. 

\subsection{Entanglement Spectrum of the Kagome Hubbard Model}
Now we focus on the CDMFT analysis of the kagome Hubbard model with $N_{\rm c}=3$. In this case, the total $4^{N_{\rm c}}=64$ eigenvalues of $\hat{\rho}_{\rm A}$ can be classified into several irreducible representations under the $U(1)\otimes SU(2)\otimes C_{3v}$ symmetry of the kagome Hubbard model. Each state can be labeled by three quantum numbers: particle number $N_p=0-6$, the $z$ component of total spin, $s_z$, and the helicity $h=0, \pm 1$. The helicity is defined from the response of the state to the 120 degrees rotation $\hat{L}$, in terms of the center of triangle, as $\hat{L}|\psi\rangle = \omega^h|\psi\rangle$, with $\omega = e^{i\frac{2}{3}\pi}$. 

In the $N_p=3$ sector, the following states are of particular interest: 
the doublon state, $|D\rangle = \frac{1}{\sqrt{3}}(|\uparrow\downarrow, \uparrow, 0\rangle + |0, \uparrow\downarrow, \uparrow\rangle + |\uparrow, 0, \uparrow\downarrow\rangle)$, the spin-polarized state, $|S\rangle = |\uparrow,\uparrow,\uparrow\rangle$, and the chiral state, $|K\rangle = \frac{1}{\sqrt{3}}(|\downarrow, \uparrow, \uparrow\rangle + \omega|\uparrow, \downarrow, \uparrow\rangle + \tilde{\omega}|\uparrow, \uparrow, \downarrow\rangle)\ (\tilde{\omega}=\omega^2)$. 
See Fig.~\ref{schematics}(b). 
We denote the eigenvalues corresponding to these states as $\rho_D$, $\rho_S$, and $\rho_K$, respectively~\cite{mixing}. 
Among these states, the chiral states are of special interest. $|K\rangle$ and its time-reversal and reflection conjugate states constitute 
fourfold degeneracy, labeled by $s_z=\pm1/2$ and $h=\pm1$.

\section{Results}
\label{results}
In this section, we will introduce our results on the entanglement spectrum. In section~\ref{heavy}, we will
introduce our previous results on the heavy-fermion behavior, and associate it with the temperature 
dependence of
entanglement spectrum at 
the $N_p=3$ sector. In section~\ref{charge}, we will show the entanglement spectrum
at 
$N_p \neq 3$ sectors, and discuss how the change of electron coherency is reflected in these spectra.

\subsection{Heavy-Fermion Behavior}
\label{heavy}
\begin{figure}[t]
\begin{center}
\includegraphics[width=0.85\textwidth]{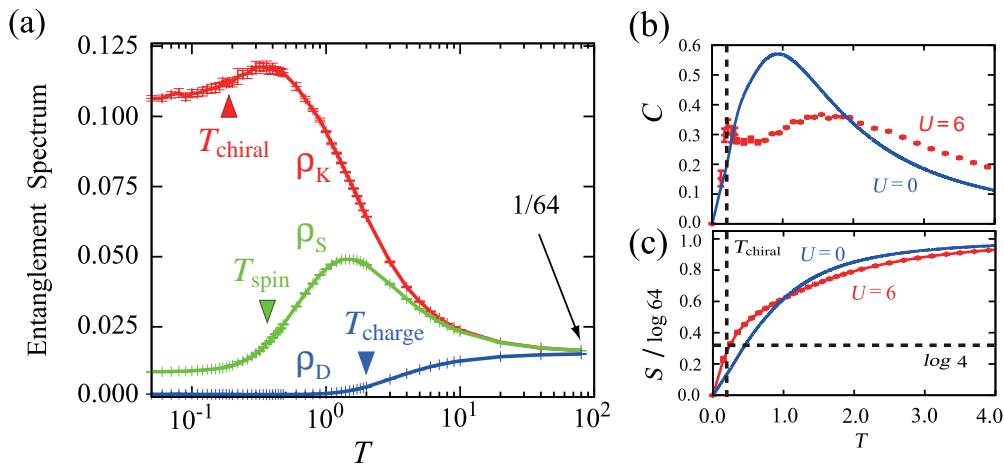}
\end{center}
\caption{\label{entanglement} 
(color online). (a) The entanglement spectrum at the $N_p=3$ sector for 
the kagome Hubbard model 
at half filling for $U=6$. $T_{\rm charge}\simeq2.0$, $T_{\rm spin}\simeq0.37$ and $T_{\rm chiral}\simeq0.18$ are the characteristic temperature scales of charge, spin, and chiral degrees of freedom,
which are determined from the inflection point of corresponding eigenvalues. (b) The temperature dependence of specific heat at $U=0$ and $6$. (c) The temperature dependence of entropy per triangular cluster, $S$, 
normalized by its value 
in the high-temperature limit, $\log 64$.}
\end{figure}

First, we summarize our results on the entanglement spectrum at $N_p=3$ sector, together with the thermodynamic behavior of this system,
which were previously discussed in ref.~\cite{Udagawa10}. In Fig.~\ref{entanglement}(a), we plot the temperature dependence of $\rho_D$, $\rho_S$, and $\rho_K$ at $U=6$,
which is slightly smaller than the critical value of $U$ 
for MIT, $U_c\sim8.3$~\cite{Ohashi06}.

At high temperature $T\sim100$, all the eigenvalues take almost the same value $\sim1/64$: 
all the 64 states are equally probable within the 
three-site cluster 
in the high-temperature limit.
As decreasing temperature, $\rho_D$ decreases at $T=T_{\rm charge}\simeq2.0$, while the other states continue to show gradual increase.
As lowering temperature further, $\rho_S$ turns to decrease at $T_{\rm spin}\simeq0.37$.
Meanwhile, $\rho_K$ continues to show enhancement even below $T_{\rm spin}$, until
another temperature scale 
$T_{\rm chiral}\simeq0.18$ is reached.
These three energy scales are defined from the inflection points of the corresponding eigenvalues. 

The two energy scales $T_{\rm charge}$ and $T_{\rm spin}$ can be basically understood from the energetics of
isolated triangle: $T_{\rm charge}$ is associated with the on-site repulsion 
$U$, and $T_{\rm spin}$ is related to
the energy scale of spin sector, $J=4t^2/U$~\cite{Udagawa10}. These states are suppressed, if temperature is lowered below $T_{\rm charge}$ or $T_{\rm spin}$. 
In other words, below $T_{\rm spin}$, the chiral states are stabilized at each triangles, and loosely interact with each other,
storing large amount of entropy $\sim\log4$ per triangle, arising from the fourfold degeneracy of the chiral state.

On the other hand, $T_{\rm chiral}$ cannot be associated with the energy scale of an isolated triangle.
This temperature rather indicates the energy scale of electron coherency, i.e., the system gains the kinetic energy from
electron motion between triangles below $T_{\rm chiral}$. Enhanced electron coherency leads to the formation of Fermi liquid state
below $T_{\rm chiral}$. This metallic state shows particularly large quasiparticle mass, since the large entropy associated
with the chiral states is now bequeathed to itinerant electrons, which leads to the steep slope of the specific heat as
shown in Fig.~\ref{entanglement}(b). In fact, the entropy takes $\sim\log4$ at $T_{\rm chiral}$ as shown in Fig.~\ref{entanglement}(c).

\subsection{Charge sectors}
\begin{figure}[t]
\begin{center}
\includegraphics[width=0.85\textwidth]{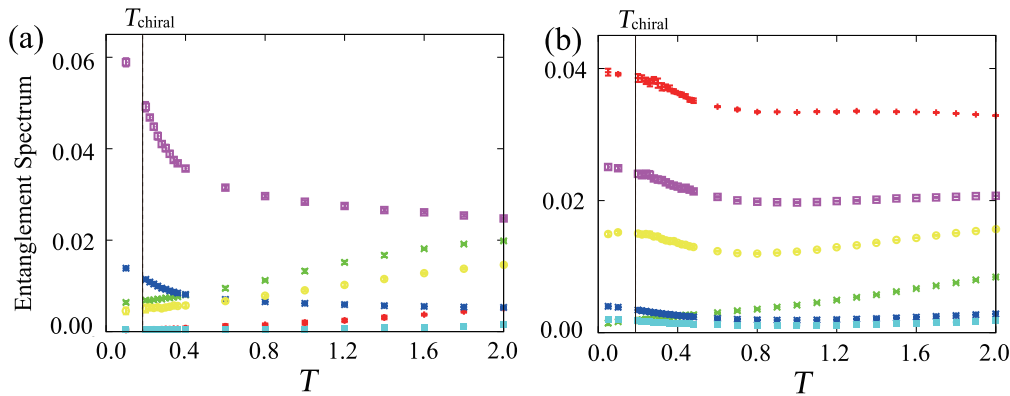}
\end{center}
\caption{\label{differentchargesectors} 
(color online). Temperature dependence of the entanglement spectra at (a) $N_p=2$ and (b) $N_p=4$ sectors. $T_{\rm chiral}$ is shown
by the vertical dashed line. The plotted 
data correspond to the states labeled by the set of quantum numbers: 
in (a), $(s_z, h, d)=(0,0,0)$, $(1,1,0)$, $(0,1,0)$, $(0,0,1)$, $(1,0,0)$, and $(0,1,1)$ 
from 
top to bottom at $T=1.6$, while in 
(b), 
$(s_z, h, d)=(1,0,1)$, $(0,0,1)$, $(0,1,1)$, $(1,1,1)$, $(0,0,2)$, and $(0,1,1)$ 
from top to bottom.}
\end{figure}

The development of electron coherency at $T
\sim T_{\rm chiral}$ is reflected in the entanglement spectra at various particle number sectors.
The electron itinerancy enhances charge fluctuations, i.e., each triangle becomes more likely to have particle numbers different from the average value, $N_p\not=3$. In Fig.~\ref{differentchargesectors}, we plot the entanglement spectra at $N_p=2$ and $4$ sectors. We here show the eigenvalues of the states labeled by the $z$ component of total spin $s_z$, helicity $h$, and doublon number $d$. While the states with different numbers of doublon mix with each other, the mixing can be ignored at low temperatures $T\ll T_{\rm charge}$. 

\label{charge}

Figure~\ref{differentchargesectors}(a) shows the temperature dependence of entanglement spectra at $N_p=2$.
The dominant contribution comes from the state with $(s_z, h, d)=(0,0,0)$. This state keeps the largest eigenvalue
of reduced density matrix from higher temperature, 
and 
starts to develop at the temperature where $\rho_K$ becomes maximum, with showing a steep enhancement at $T \sim T_{\rm chiral}$.

Figure~\ref{differentchargesectors}(b) gives the behavior of entanglement spectra at $N_p=4$.
In this sector, the states with a finite total spin have the largest eigenvalue, in contrast to $N_p=2$.
Indeed, the states with $s_z=\pm1$ 
have the lowest eigenenergy within the states at $N_p=4$ in the three-site Hubbard model,
which explains why 
they keep dominant. 
These states, together with several subdominant states also 
exhibit a slight enhancement at $T
\sim T_{\rm chiral}$.

These temperature dependences of the entanglement spectra around $T=T_{\rm chiral}$ give a clue to understanding the nature of metallic states at low temperatures.
While the eigenvalue corresponding to the chiral states, $\rho_K$, shows a slight reduction at $T
\sim T_{\rm chiral}$ as shown in Fig.~\ref{entanglement}(a),
it still keeps dominant among the 64 possible states; 
locally on each triangle, the chiral state is most likely realized even below $T
\sim T_{\rm chiral}$. 
Hence, it is reasonable to consider that electron coherency can be achieved by the enhancement of electron hopping in the sea of chiral states.

\begin{figure}[t]
\begin{center}
\includegraphics[width=0.8\textwidth]{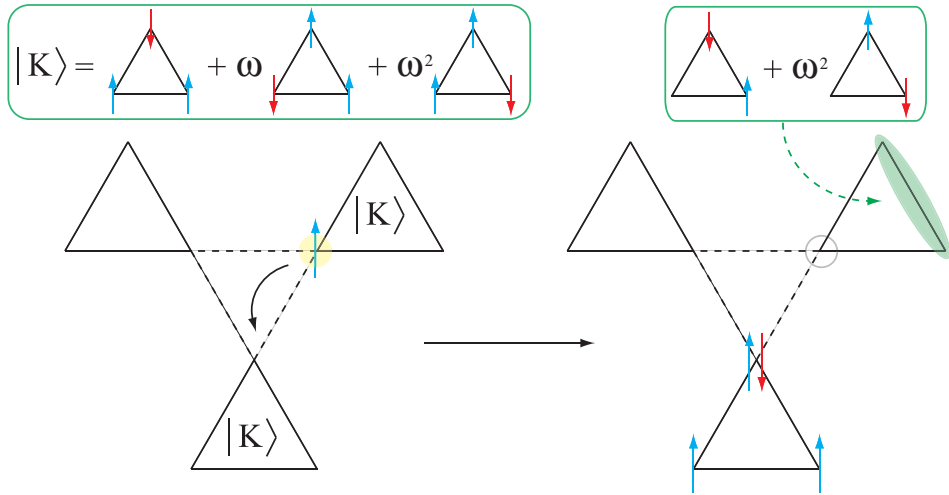}
\end{center}
\caption{\label{kagometransfer} 
(color online). A schematic picture of electron transfer in the sea of chiral states. 
Suppose all the triangles are in the state $|K\rangle$ shown in the top left.
If one spin-up electron is transferred from the left bottom
site of one triangle to the top site of the neighboring triangle, the first triangle will be left with one up-spin and one down-spin electrons with $s_z=0$, while the second triangle will acquire total spin $s_z=1$.}
\end{figure}

Suppose each triangle takes one of the fourfold degenerate chiral state. If an electron is moved from one triangle to neighboring one,
then the first triangle is left with two electrons, while the latter ends up in having four electrons. 
The enhanced eigenvalue of $s_z=0$ state in the $N_p=2$ sector suggests that the spin-up (spin-down) electron is more likely to hop from
the chiral state with $s_z=1/2$ ($-1/2$). Meanwhile among the states in the $N_p=4$ sector, $s_z=\pm 1$ states are dominant.
This means that the spin-up (spin-down) electron tends to hop to the chiral states with $s_z=1/2$ $(-1/2)$ (Fig.~\ref{kagometransfer}).
Therefore, for the single-electron hopping process to create $s_z=0$ state with $N_p=2$ and $s_z=1 (-1)$ state with $N_p=4$ simultaneously,
the chiral states have to have a pair of $s_z=1/2 (-1/2)$ states at neighboring triangles. In other words, the entanglement spectra
imply that short-range ferromagnetic correlations 
are developed between neighboring triangles in the correlated metallic phase.

\label{coherency}

\section{Discussion and Summary}
\label{sec:summary}
In this letter, we 
have studied the entanglement spectra defined from the thermal density matrix in the Hubbard model at half filling on the kagome lattice.
The cluster dynamical mean-field theory provides a convenient tool to analyze this quantity through the whole temperature range.
Indeed, we have clarified that the entanglement spectra for a three-site cluster exhibit nontrivial temperature dependence.

At the $N_p=3$ sector, the clear separation of energy scales between charge, spin, and chiral degrees of freedom is obtained.
In particular, the characteristic temperature scale of chiral degrees of freedom, $T_{\rm chiral}$, gives a good measure
to the enhancement of electron coherency. In fact, the Fermi liquid state with large quasiparticle mass is formed below this temperature,
as clearly shown in the analysis of specific heat and entropy.

Moreover, the developed electron coherency affects the entanglement spectra at $N_p=2$ and $4$ sectors.
The dominant states at both sectors show enhancement at $T\sim T_{\rm chiral}$, suggesting 
enhanced charge fluctuations due to the electron itinerancy. The growth of these states implies 
short-range inter-triangle ferromagnetic correlations in 
the
low-temperature metallic state. While the complete characterization of the correlated metallic state is left for a future task,
the entanglement spectra will serve as a powerful tool for this purpose.

\section*{Acknowledgements}
This work was supported by Grants-in-Aid for Scientific Research
(No. 24340076, 26400339, and 24740221), the Strategic Programs for Innovative Research (SPIRE),
MEXT, and the Computational Materials Science Initiative (CMSI), Japan.


\section*{References} 
\begin{thebibliography}{10}
\bibitem{Georges96} A. Georges et al., Rev. Mod. Phys. {\bf 68}, 13 (1996).
%
\bibitem{Georges93} A. Georges and W. Krauth, Phys. Rev. B {\bf 48}, 7167 (1993).
%
\bibitem{Kotliar01} G. Kotliar et al., Phys. Rev. Lett. {\bf 87}, 186401 (2001).
%
\bibitem{Gull08} E. Gull et al., Europhys. Lett. {\bf 82}, 57003 (2008).
%
\bibitem{Udagawa10} M. Udagawa and Y. Motome, Phys. Rev. Lett. 
{\bf 104}, 106409 (2010).
%
\bibitem{Imai03} Y. Imai, N. Kawakami, and H. Tsunetsugu, Phys. Rev. B {\bf 68}, 195103 (2003).
%
\bibitem{Bulut05} N. Bulut, W. Koshibae, and S. Maekawa, Phys. Rev. Lett. {\bf 95}, 037001 (2005).
%
\bibitem{Ohashi06} T. Ohashi, N. Kawakami, and H. Tsunetsugu, Phys. Rev. Lett. {\bf 97}, 066401 (2006).
%
\bibitem{Bernhard07} B. H. Bernhard, B. Canals, and C. Lacroix, J. Phys.: Condens. Matter {\bf 19}, 145258 (2007).
%
\bibitem{Furukawa10} Y. Furukawa, T. Ohashi, Y. Koyama, and N. Kawakami, Phys. Rev. B, {\bf 82}, 161101 (2010).
%
\bibitem{Udagawa10_2} M. Udagawa and Y. Motome, J. Phys.: Conf. Ser. {\bf 200}, 012214 (2010). 
%
\bibitem{mixing} Although $|D\rangle$ has finite off-diagonal components with the state, $\frac{1}{\sqrt{3}}(|\downarrow, \uparrow, \uparrow\rangle + |\uparrow, \downarrow, \uparrow\rangle + |\uparrow, \uparrow, \downarrow\rangle)$, we ignore it and focus on the diagonal component,
to characterize the energy scale associated with the charge degree 
of freedom.
\end{thebibliography}
\end{document}